\documentclass[pdflatex,sn-nature]{sn-jnl}% Style for submissions to Nature Portfolio journals
\usepackage{graphicx}%
\usepackage{multirow}%
\usepackage{amsmath,amssymb,amsfonts}%
\usepackage{amsthm}%
\usepackage{mathrsfs}%
\usepackage[title]{appendix}%
\usepackage{xcolor}%
\usepackage{textcomp}%
\usepackage{manyfoot}%
\usepackage{booktabs}%
\usepackage{algorithm}%
\usepackage{algorithmicx}%
\usepackage{algpseudocode}%
\usepackage{listings}%

\usepackage{parskip}
\usepackage{gensymb}
\usepackage{ragged2e}
\usepackage{lineno}

\usepackage[maxnames=99,sorting=none, style=ieee, citestyle=numeric-comp, backend=biber]{biblatex}
\begin{document}

\newgeometry{top=0.75in,left=1.25in,right=1.25in}
%\linenumbers

\title[Article Title]{\textbf{Ultrafast photocurrent detection reveals that device efficiency is dominated by ultrafast exciton dissociation not exciton diffusion}}

%%=============================================================%%
%% GivenName	-> \fnm{Joergen W.}
%% Particle	-> \spfx{van der} -> surname prefix
%% FamilyName	-> \sur{Ploeg}
%% Suffix	-> \sfx{IV}
%% \author*[1,2]{\fnm{Joergen W.} \spfx{van der} \sur{Ploeg} 
%%  \sfx{IV}}\email{iauthor@gmail.com}
%%=============================================================%%

\author[1]{\fnm{Zachary M.} \sur{Faitz}}\email{zanni@chem.wisc.edu}

\author[2]{\fnm{Christopher J.} \sur{Blackwell}}\email{cblackwell@wisc.edu}

\author[1]{\fnm{Dasol} \sur{Im}}\email{dim4@wisc.edu}

\author[2]{\fnm{Abitha} \sur{Dhavamani}}\email{adhavamani4@wisc.edu}

\author[2]{\fnm{Michael} \sur{Arnold}}\email{michael.arnold@wisc.edu}

\author*[1]{\fnm{Martin T.} \sur{Zanni}}\email{zanni@chem.wisc.edu}

\affil[1]{\orgdiv{Department of Chemistry}, \orgname{University of Wisconsin - Madison}, \orgaddress{\street{1101 University Ave.}, \city{Madison}, \postcode{53711}, \state{Wisconsin}, \country{United States}}}

\affil[2]{\orgdiv{Department of Materials Science and Engineering}, \orgname{University of Wisconsin - Madison}, \orgaddress{\street{1101 University Ave.}, \city{Madison}, \postcode{53711}, \state{Wisconsin}, \country{United States}}}

\abstract{Excitons diffusing to a charge-separating interface is a necessary step to convert energy into current in next-generation photovoltaics. In this report, made possible by a new ultrafast spectrometer design, we compare exciton dynamics measured using both photoabsorption- and photocurrent-detected transient and 2D spectroscopies. For a device with semiconducting carbon nanotubes as the exciton transport material, we find that photoabsorption detection greatly overestimates the importance of long-lived excitons for device performance. Excitons diffuse and transfer between nanotubes for several picoseconds, but the large majority of photocurrent is created within 30fs by excitons that diffuse little to the C\textsubscript{60} electron transfer material. These results change our understanding of the material features most important for these photovoltaics. Photoabsorption detection measures all excitons, but not all photogenerated excitons generate current. To understand device efficiency, this study points to the necessity for directly measuring the exciton dynamics responsible for photocurrent. 
}


%%\pacs[JEL Classification]{D8, H51}

%%\pacs[MSC Classification]{35A01, 65L10, 65L12, 65L20, 65L70}

\maketitle
\newpage
\newgeometry{top=1in,bottom=2in,left=1.5in,right=1.5in}

\section{Introduction}\label{sec1}

Many next-generation solar cells generate photocurrent by harvesting excitons.\cite{Gunes_2007,Rao_2017,Arnold_2013,Gregg_2003,Feron_2012} In this device design, photons generate electron-hole pairs (excitons) which are separated into free carriers at the interface of electron and hole transfer materials. The charges flow through their respective domains to the electrodes, creating current. To generate current, excitons must reach an interface before recombining. Thus, exciton dynamics and device efficiency are intricately linked to materials morphology and microstructure in this class of next-generation photovoltaics.\cite{Lee_2022,Shea_2018,Zeng_2023,Hu_2018}

\par

Ideally, one knows both the pathways for excitons to reach the heterojunction and those that end in recombination. Transient absorption, also known as pump-probe spectroscopy, is the most common method for measuring these pathways.\cite{Hu_2018,Jailaubekov_2013,Ohkita_2011,Gross_2023,Wan_2015} In this technique, a pump pulse creates excitons which are monitored by the change in absorbance of a probe pulse. It is often applied to films with the assumption that the properties of the photoactive material are unchanged in a device. Absorption detection measures all excitons within the probe’s bandwidth, regardless of the exciton’s fate. In contrast, photocurrent detection does not report on recombined or trapped excitons, but only measures excitons that ultimately generate current.\cite{Bakulin_2016} Even the best next-generation photovoltaics do not create current from every photon absorbed, so discerning relevant \textit{in situ} processes is difficult with absorption detection alone. Ideally, one would use both absorption and photocurrent detection to obtain a comprehensive understanding of current-generating and non-current-generating pathways. 

\par

Photocurrent detection is commonly used to evaluate devices but infrequently used in ultrafast pump-probe and 2D experiments that measure exciton dynamics.\cite{Nardin_2013,Bolzonello_2021,Karki_2014} Ultrafast photocurrent detection suffers from an “incoherent” background that is created by exciton-exciton and/or exciton-charge interactions that occur after the probe pulse and can be orders-of-magnitude larger than the pump-probe signals.\cite{Bargigia_2022,Gregoire_2018,Bruschi_2023,Bolzonello_2023,Javed_2024} Recently, two methods for removing the incoherent signals have been reported.\cite{Faitz_2024,Charvatova_2025} In this manuscript, we utilize one of these techniques, a new polarization pulse sequence that gives background-free, photocurrent-detected spectra in isotropically ordered samples.\cite{Faitz_2024}

\par

In this work, we utilize films and photovoltaic devices made from single-walled, semiconducting carbon nanotubes (CNTs) that act as the light-absorbing and exciton transport material.\cite{Saito_1995} CNTs are promising materials for photovoltaics because they are extremely strong absorbers,  their bandgap is easily tuned from the visible to the near-IR based on their diameter (referred to by chiral index (n,m)), and they are chemically stable and solution-processable.\cite{Yi_2007,Arnold_2013,Mistry_2013} Excitons are highly stable in CNTs because their one dimensional nature reduces dielectric screening. Additionally, their wire-like geometry and fast exciton diffusivity allows for rapid exciton transport ($\>>$100nm along their length). Their narrow linewidths ($\sim$30nm) make them excellent materials for spectroscopic study because exciton and charge transfer can be monitored between different bandgap CNTs. Indeed, we know from transient absorption spectroscopy that excitons exist and transfer in these films for several picoseconds, which has guided the design of CNT photovoltaics.\cite{Flach_2020} Here, we report that transient absorption spectroscopy vastly overestimates the importance of inter-tube exciton hopping and diffusion for device performance because the photocurrent is created almost entirely by excitons that require little-to-no diffusion. These realizations are leading us to reevaluate the photophysics that dictate CNT photovoltaic device function.
 
\bigskip
\section{Results}\label{sec2}

\begin{figure}[h]
\centering
\includegraphics[width=0.9\textwidth]{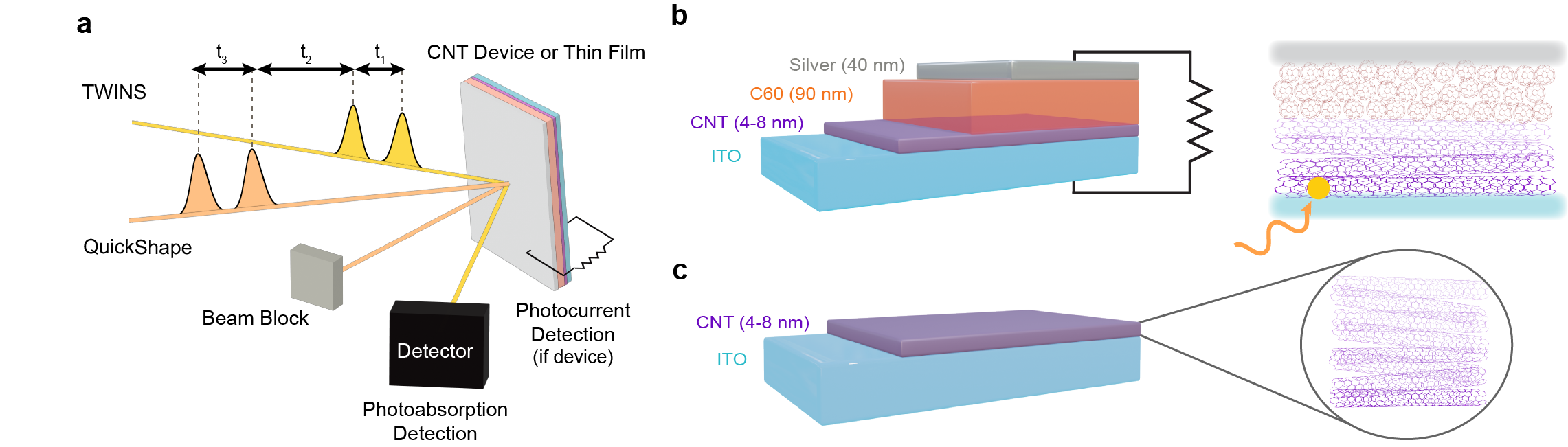}
\caption{ Instrument and sample schematics. (a) Schematic of instrument geometry used to measure simultaneous 2D photocurrent and 2D absorption spectra of CNT photovoltaic devices. The device can be replaced with a thin film, in which case only 2D absorption is measured. (b) A simplified diagram of a (6,5) CNT photovoltaic device. A detailed description of the device is found in Methods. (c) A diagram of the (6,5) thin film used in 2D absorption-only measurements.}\label{fig1}
\end{figure}

In this work, we measure three devices and one thin film. The devices use a planar heterojunction architecture with layers of CNT and C\textsubscript{60} between an ITO anode and a Ag cathode (Fig. \ref{fig1}; Methods).\cite{Bindl_2013} The three devices differ in their composition of CNT, being either solely (6,5), solely (7,5), or equal portions of (6,5) and (7,5) CNTs that are mixed before deposition. The film is solely (6,5) CNTs deposited on ITO. 

\par

Photons are absorbed by the CNT to create excitons (Fig. \ref{fig1}b, yellow dot). Excitons diffuse along the lengths of CNTs and hop between CNTs.\cite{Crochet_2012,Birkmeier_2022,Mehlenbacher_2015} Recombination occurs when the exciton reaches the end of a tube or a trap state, which exists about every 140 nm.\cite{Wang_2017} If the exciton encounters C\textsubscript{60}, it can dissociate into an electron and hole. The holes travel back through the CNTs to the anode while electrons travel through C\textsubscript{60} to the cathode, resulting in current. Understanding these processes is key to utilizing CNTs in photovoltaics. 

\par

These experiments are made possible by a new ultrafast spectrometer that can simultaneously measure photoabsorption- and photocurrent-detected spectra free from incoherent background (Fig. \ref{fig1}; Methods).\cite{Faitz_2024} As explained in the Analysis section, all excitons and holes are observed with absorption detection (with holes causing half the typical bleach signal; see SI) whereas only excitons and holes that ultimately produce current are measured in photocurrent detection.  As we show, a comparison of photoabsorption and photocurrent detection provides insight into exciton processes that are important for photovoltaic microstructure design. 

\bigskip
\subsection{Comparison of thin film and device dynamics}\label{subsec2}

\begin{figure}[h]
\centering
\includegraphics[width=0.9\textwidth]{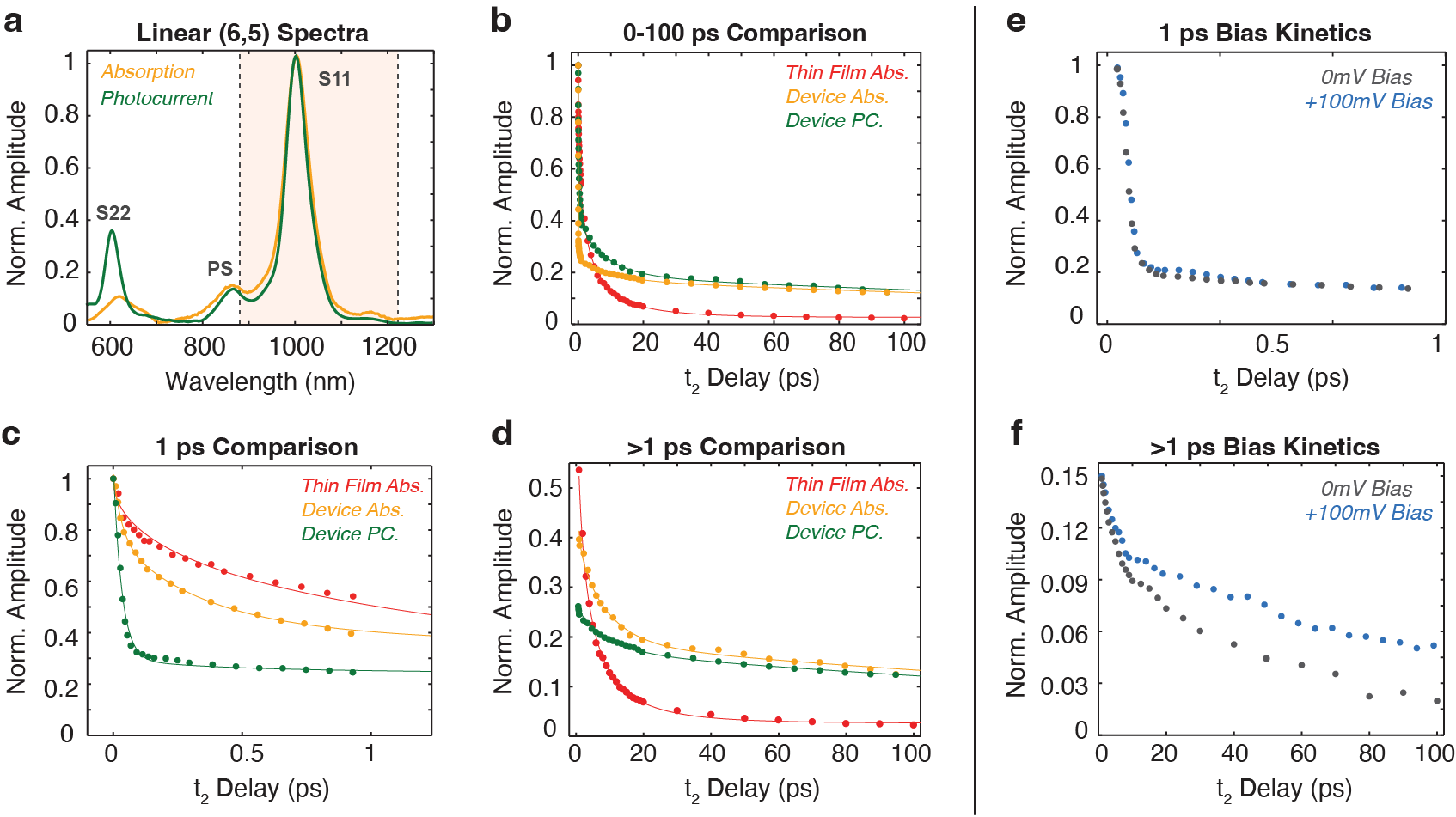}
\caption{ Comparison of kinetics for a (6,5) film and device measured with pump-probe photoabsorption and photocurrent detection. (a) Overlayed linear absorption (yellow) and photocurrent (green) spectra for a photovoltaic. Three transitions are labeled: the S\textsubscript{11} at 1000nm, the phonon sideband (ps) at 860nm, and the S\textsubscript{22} at 580nm. The shaded pink region represents the spectral width of the laser pulse. (b-c) Normalized transient absorption and photocurrent kinetics taken at 1000nm (S\textsubscript{11}) for a thin film (red), an absorption-detected device (yellow), and a photocurrent-detected device (green). The solid lines in each plot are the fits described in the Analysis and Discussion section below. (b) Plots the full kinetics out to 100ps, (c) plots the first 1 ps of the kinetics data, and (d) plots t\textsubscript{2} delays  $\>>$1ps. (e,f) Kinetics traces of unbiased (blue) and biased (grey) transient photocurrent data. (e) Plots the first 1ps of data, and (f) plots the t\textsubscript{2} delays $\>>$1ps. Two different photovoltaic devices were used to collect the data in (b-d) vs. (e,f), which is why their amplitudes are slightly different (see SI for examples of device variation).}\label{fig2}
\end{figure}

We start by presenting data collected using films and devices made from (6,5) CNTs. Shown in Fig. \ref{fig2}a are the linear absorption and photocurrent spectra of a (6,5) device. The internal quantum efficiency of our devices is $\sim$50\%, meaning half of the excitons do not generate current. The (6,5) CNTs primarily absorb at 1000nm with a phonon sideband absorption at 860nm. These are linear spectra that do not report on exciton dynamics, so one cannot use them to determine if excitons dissociate to charges immediately or first transfer between nanotubes to reach the C\textsubscript{60} heterojunction. 

\par

Shown in Fig. \ref{fig2}b-d are transient absorption kinetics for a (6,5) CNT film as well as kinetics for a (6,5) CNT device measured with both transient absorption and transient photocurrent at 1000nm. It is apparent that each measurement exhibits different kinetics on multiple timescales. On the sub-picosecond timescale (Fig. \ref{fig2}c), the slowest kinetics are observed in the thin film, and the fastest kinetics are observed in the photocurrent-detected device. Between 1 and 100ps (Fig. \ref{fig2}d), the signal for the thin film decays to baseline while the device signals still have $\sim$10\% of their initial intensity. Measurements for a device made from (7,5) CNTs are nearly identical to the (6,5) device (see SI), which is expected because the excitons of both films will dissociate upon contact with C\textsubscript{60}.\cite{Wang_2019} No detailed analysis is needed to conclude that the kinetics in these three measurements are so different that either the photophysics of the films are altered in a device or that photoabsorption and photocurrent detection report on different photophysics.

\par 

When excitons dissociate, they generate charges, so device measurements should contain spectroscopic signatures of both excitons and holes.\cite{Bolzonello_2021,Hinrichsen_2020} To distinguish between the two, we repeated the measurements with an applied bias since only charges are influenced by electric fields. Shown in Fig. \ref{fig2}e,f are kinetics measured with and without a +100mV forward bias. The kinetic traces are normalized at t\textsubscript{2}=0fs to account for the decrease in charge collection efficiency under forward bias. Before 1ps, no significant difference in kinetics is observed with the bias. In contrast, for times after $\sim$5ps, the signal decays more slowly upon application of the +100mV bias. Slower kinetics are consistent with a positive bias, which should hinder hole transport to the electrode. Thus, the sub-ps timescales are likely dominated by exciton dynamics while longer timescales contain hole dynamics.   

\bigskip
\subsection{Exciton and hole transfer between carbon nanotubes}\label{subsec2}

The above experiments on single-chirality CNTs establish the fundamental timescales for the photophysics in these systems, but they do not resolve exciton or charge transfer between CNTs. To observe these events, we utilized 2D spectroscopy and devices with a photoabsorbing layer made from a mixture of (6,5) and (7,5) CNTs. Shown in Fig. 3a are linear photocurrent spectra of devices made from solely (6,5) or (7,5) nanotubes, overlayed on the photocurrent spectrum for the mixed (6,5)/(7,5) device. (7,5) nanotubes absorb at 1050nm and are spectroscopically resolvable from (6,5) nanotubes. Cross peaks in the 2D spectra will be created by either exciton or hole transfer, enabling experiments that monitor inter-tube transfer.

\begin{figure}[h]
\centering
\includegraphics[width=0.9\textwidth]{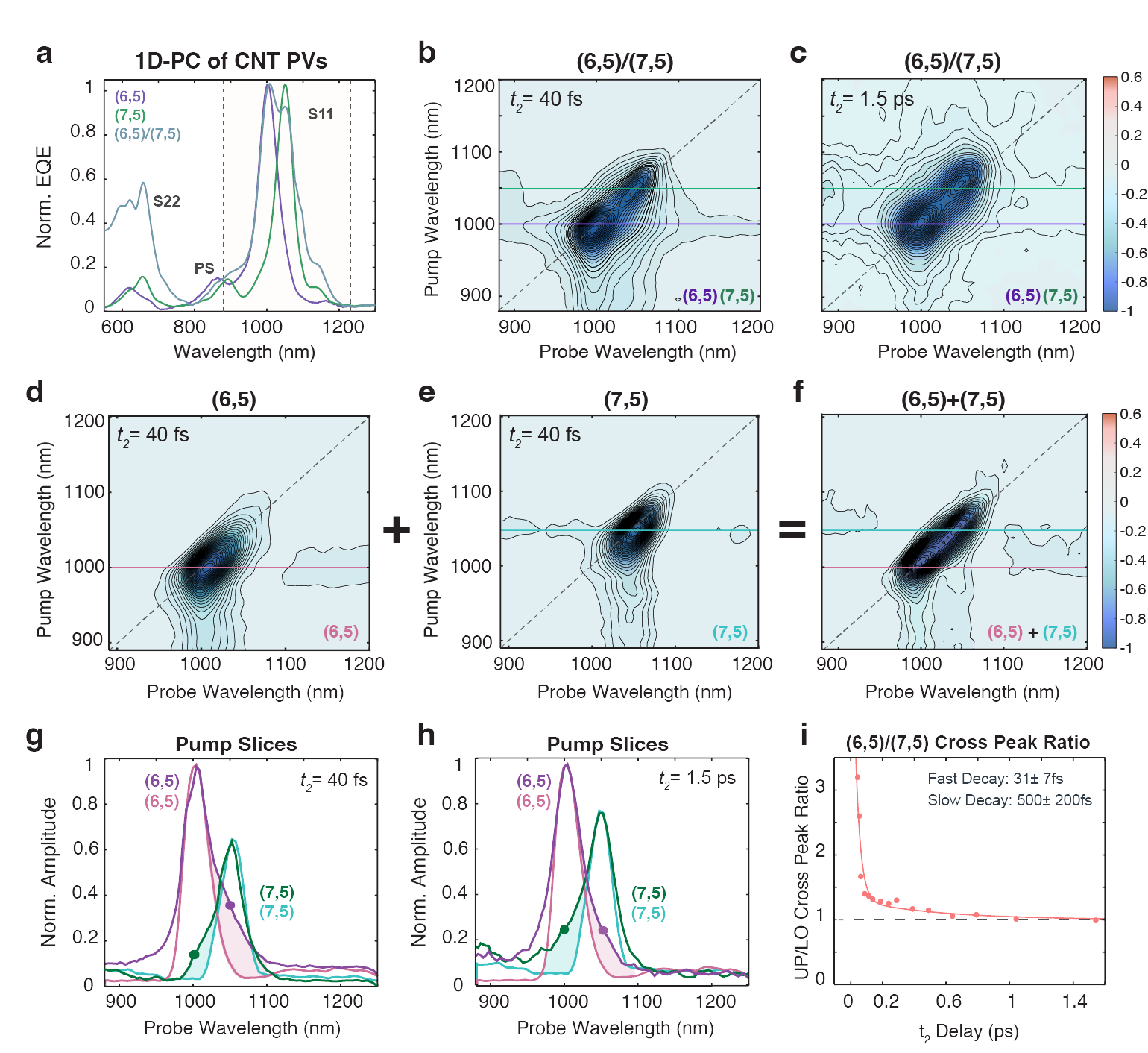}
\caption{ 1D and 2D photocurrent-detected spectra for devices of (6,5), (7,5), and mixed (6,5)/(7,5) chiralities. (a) Overlapped and normalized linear photocurrent spectra of (6,5) (purple), (7,5) (green), and (6,5)/(7,5) (blue). (b,c) Normalized 2D photocurrent spectra from a mixed (6,5)/(7,5) photovoltaic device at 40fs and 1500fs, respectively. (d-f) Depicts the process of mathematically adding single chirality spectra: spectra of (d) (6,5) and (e) (7,5)  single chirality photovoltaic devices at 40fs are mathematically added to create (f) a (6,5)+(7,5) 2D photocurrent spectrum. (g,h) Pump slices at 1000nm (purple and pink) and at 1050nm (green and teal) from both the (6,5)/(7,5) mixed device and the mathematically added (6,5)+(7,5) device. Both plots are marked with purple and green dots which represent the intensity of the upper and lower cross peaks, respectively. Also, in both plots, the region between the mixed (6,5)/(7,5) slices and the mathematically added (6,5)+(7,5) pump slices are shaded. (i) The ratio of the lower cross peak to the upper cross peak as a function of t\textsubscript{2} delay calculated using the intensity at the frequency of the colored points in (g) and (h). The data is fit to a biexponential decay. }\label{fig3}
\end{figure}

\par

Shown in Fig. \ref{fig3}b,c are the 2D photocurrent spectra of the mixed (6,5)/(7,5) device at t\textsubscript{2}=40fs and 1500fs, respectively. Two diagonal peaks are observed, corresponding to the bandgaps of the (6,5) and (7,5) nanotubes. There also exist bulges in the off-diagonal portions of the spectra where cross-peaks are expected. To assess these potential cross-peaks, we present 2D photocurrent spectra in Fig. \ref{fig3}d,e for devices made of (6,5) and (7,5) CNTs, respectively. These spectra contain one predominant peak corresponding to the bandgap of either the (6,5) or (7,5) CNTs. A fictional 2D spectrum is shown in Fig. 3f, created by mathematically summing the (6,5) and (7,5) 2D spectra (Fig. \ref{fig3}d,e). The summed spectrum cannot include energy or charge transfer, so it lacks cross-peaks and confirms the features in Fig. \ref{fig3}b,c are created by either exciton transfer, charge transfer, or both.

\par

We also find the cross-peak intensities are not equal on the two halves of the spectrum at sub-picosecond delays. Shown in Fig. \ref{fig3}g,h are horizontal cuts at 1000 and 1050nm through the 2D photocurrent spectra collected at t\textsubscript{2}=40 and 1500fs, respectively. The cuts are superimposed on the corresponding cuts through the mathematically summed (6,5)+(7,5) spectrum. Notice the cross-peaks, which appear as shoulders on the sides of the larger diagonal peaks (shaded teal or pink), have unequal amplitude at t\textsubscript{2}=40fs. The cross-peak below the diagonal, corresponding to downhill energy transfer, is more intense initially, but by t\textsubscript{2} $\approx$1000fs, the asymmetry is gone (Fig. \ref{fig3}h,i). As explained in the Analysis below, both exciton and hole transfer contribute to the cross-peaks. 

\bigskip
\section{Analysis and Discussion}\label{sec3}

The experiments presented above include absorption detection of a film, absorption detection of devices, and photocurrent detection of devices. Upon inspection, the kinetics look very different (Fig. \ref{fig2}), suggesting unrelated photophysical processes. But as we show below, the data can be fit by a unified kinetic model that includes exciton kinetics, exciton dissociation, and hole kinetics. We find that all three measurements are described by the same timescales, but with each photophysical process having a different amplitude depending on the sample and type of detection. 

\bigskip
\subsection{Exciton versus hole timescales}\label{subsec3}

\begin{figure}[h]
\centering
\includegraphics[width=0.9\textwidth]{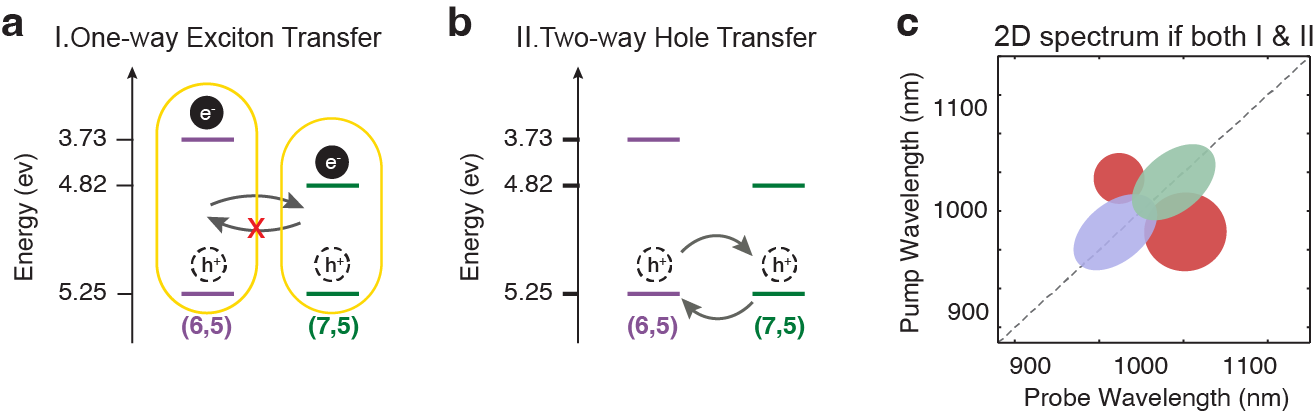}
\caption{ A diagram depicting the origin of asymmetric cross peaks in 2D photocurrent spectra. (a) Energy level diagram for (6,5) and (7,5) CNTs with valance band minimum and conduction band maximum energy level depicting the downhill energy transfer of excitons. (b) Energy level diagram depicting two-way hole transfer between (6,5) and (7,5) CNTs. (c) The resulting 2D photocurrent spectrum with asymmetric cross peaks if both processes depicted in (a) and (b) are present in the system. Hole transfer can create symmetric cross peaks while exciton transfer only appears as a lower cross peak.}\label{fig4}
\end{figure}

A key aspect of interpreting these kinetics is the assignment of timescales associated with excitons versus holes. To do so, we start by interpreting the 2D photocurrent data in Fig. \ref{fig3}. Shown in Fig. \ref{fig4}a,b  are the energy levels of the valence and conduction bands at k=0 momentum for (6,5) and (7,5) CNTs. The difference in bandgap between the (6,5) and (7,5) CNTs is much larger than kT, so exciton transfer can only occur downhill in energy from (6,5) to (7,5) CNTs (Fig. \ref{fig4}a), creating a cross-peak on the lower half of the spectrum.  If the cross-peaks were solely due to exciton transfer, no cross-peak would be observed on the upper half of the 2D spectrum. The bandgaps of (6,5) and (7,5) CNTs are different because of the energies of their conduction bands, but their valence bands are nearly degenerate.\cite{Wang_2019} As a result, hole transfer can occur from (6,5) to (7,5) tubes and vice versa, creating cross-peaks on both sides of the diagonal with equal magnitude. Thus, upper and lower cross-peaks with a more intense lower cross-peak is a clear signature that both exciton and hole transfer occurs (Fig. \ref{fig5}c). With this assignment, we fit the ratio of the cross-peak amplitudes (Fig. \ref{fig3}i) to a biexponential decay, giving time constants of 31 $\pm7$fs and 500 $\pm200$fs for exciton transfer from (6,5) to (7,5) CNT.  At t\textsubscript{2}$\approx$1ps, the cross-peaks have equal intensities (Fig. \ref{fig3}c,h, until $\sim$30ps; see SI). Thus, after 1ps, excitons that have transferred between tubes no longer contribute to the photocurrent, and only hole transfer occurs between CNTs.  

\bigskip
\subsection{Unified model for a global fit of kinetics}\label{subsec3}

To fit the kinetics in Fig. \ref{fig2}, we present the following model. The recombination rate of excitons is modeled by a stretched exponential (Eqn. 1), following prior work that linked the exciton lifetime to defect densities.\cite{Flach_2020, Wang_2017} For all other photophysical processes, such as exciton dissociation and hole transfer, we assume first-order rate equations (Eqn. 2,3). 

\begin{align}
&\frac{dP_{\text{recom}}}{dt} = -\frac{1}{2a} \left(\frac{t}{a}\right)^{-1/2} P_{\text{recom}}(t)\\
&\frac{dP_{\text{dis}}}{dt} = -bP_{\text{dis}}(t) -cP_{\text{dis}}(t)\\
&\frac{dH}{dt} = bP_{\text{dis}}(t) +cP_{\text{dis}}(t) - dH(t) - eH(t)
\end{align}

\par

In the above equations, $P_{\text{recom}}$, $P_{\text{dis}}$, and $H$ represent the populations of excitons that recombine, excitons that dissociate, and holes, respectively. $a$ – $e$ are the time constants for each process. To calculate the population of excitons as a function of time, we integrate the rate equations (see SI). We find a minimum of four first-order rate equations is necessary to fit the data (two for Fig. \ref{fig2}c and two for Fig. \ref{fig2}d). We also note that according to the voltage-dependent kinetics (Fig. \ref{fig2}e,f) and the asymmetric cross-peaks (Fig. \ref{fig3}i), only holes contribute to the photocurrent after $\sim$1ps.  Thus, two of the rate equations are assigned to excitons and two to holes. Since the dissociation of an exciton is required to create a hole, these two sets of equations are coupled. The resulting amplitudes and time constants are used to simulate the kinetics (see SI), noting that not all photophysical processes contribute to each measurement. The dominant photophysical process in the films is the exciton lifetime (Eqn. 1) since only a very small percentage of excitons dissociate in films (See SI).\cite{Bindl_2013_2} In contrast, the exciton lifetime does not contribute to the photocurrent data because photocurrent is not generated from excitons that recombine. All five physical processes are included in the device transient absorption experiments because absorption is sensitive to both excitons and holes regardless of whether they contribute to photocurrent.

\par

Using this model, we globally fit the kinetic data in Fig. \ref{fig2} from 0 to 100ps by constraining the time-constants to be the same across measurements while allowing relative amplitudes to vary. The fits are excellent, and the resulting amplitudes and time-constants are given in Table \ref{tab1}. We discover that each photophysical process contributes a different amount to each of the three measurements. For example, 30fs exciton dynamics dominate the photocurrent kinetics (89\%), contribute moderately to the photoabsorption kinetics of the devices (33\%), and do not contribute to film kinetics. As a result, the kinetic traces appear to be disparate but the transient absorption and transient photocurrent data actually contain the same photophysics.  
\begin{center}
\begin{table}[h]

%\begin{tabular}{@{}llllll@{}}
\begin{tabular}{p{0.16\linewidth} | p{0.14\linewidth}  p{0.11\linewidth}  p{0.11\linewidth}  p{0.11\linewidth}  p{0.11\linewidth}}

Time Constants &  \centering 2 $\pm$0.1ps  & \centering 30 $\pm$3fs & \centering 600 $\pm$100fs & \centering 9 $\pm$1ps & \Centering 260 $\pm$20ps\\
\midrule
Film Abs.    & \centering 1.0  & \centering 0  & \centering 0 & \centering 0 &  \Centering 0  \\
Device Current    &  \centering 0   & \centering 0.89 $\pm$0.02  & \centering 0.11 $\pm$0.01 & \centering 0.27 $\pm$0.03 &  0.73 $\pm$0.03 \\
Device Abs.   &  \centering  0.21 $\pm$0.02   & \centering 0.33 $\pm$0.02  & \centering 0.46 $\pm$0.03 & \centering 0.21 $\pm$0.03 & 0.79 $\pm$0.04  \\
\midrule
\vfill Assignment   & \vfill \centering Recombination   &   \centering Adjacent   to C\textsubscript{60}  &   \centering Diffusion to C\textsubscript{60} & \centering Fast Hole Collection &  \Centering Slow Hole Collection \\

\midrule
\end{tabular}

\begin{justify}
\caption{\justifying Time constants and relative amplitude of the photophysical processes observed in the kinetics data in Fig. \ref{fig2}. Not all physical processes are present for each type of sample or mode of detection. The amplitudes for the populations of the excitons and holes are normalized separately. The reported errors in time constants and amplitudes are the standard errors calculated based on the sum of square residuals and the degrees of freedom in the fit. See SI for additional details.}\label{tab1}
\end{justify}

\end{table}
\end{center}

\bigskip
\subsection{Physical interpretation of the photophysics and deductions for device design}\label{subsec3}

\begin{figure}[h]
\centering
\includegraphics[width=0.9\textwidth]{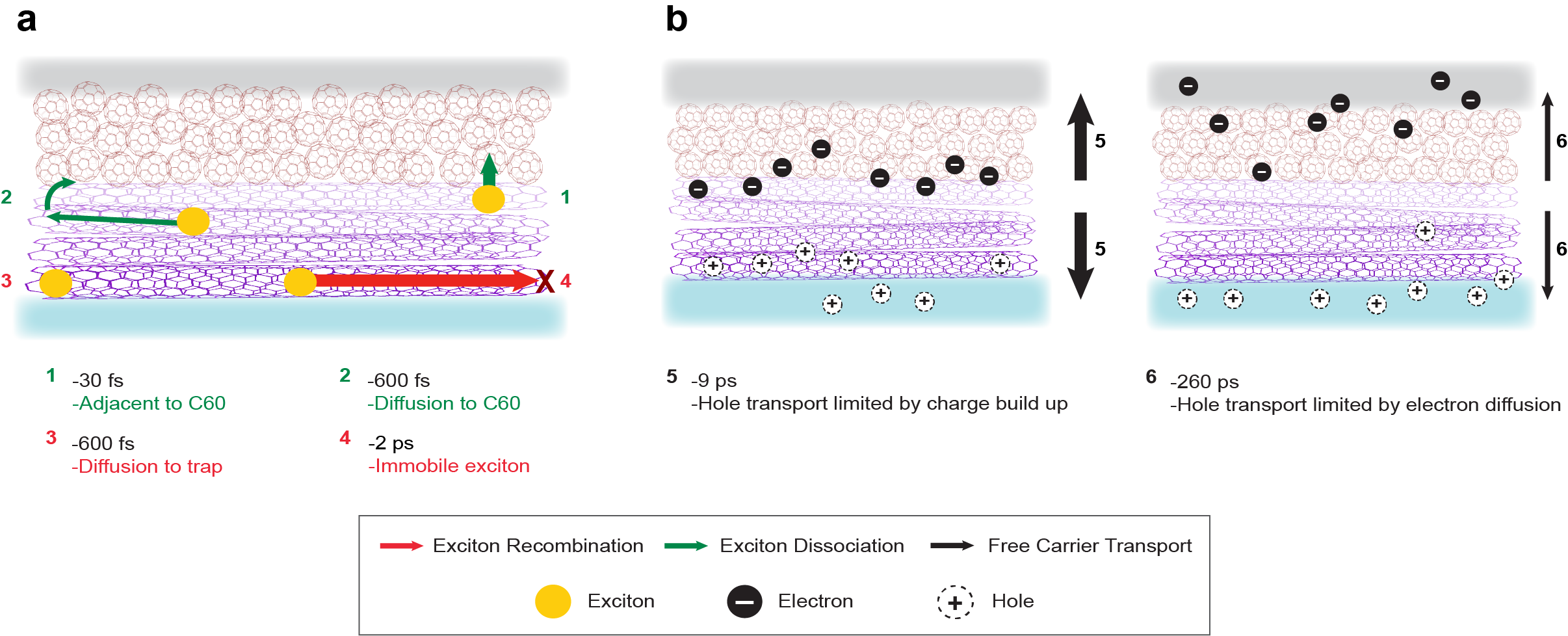}
\caption{Diagrams of the physical processes listed in Table \ref{tab1}. (a) Exciton-related processes. (b) Carrier related processes. The color of the arrows corresponds to exciton recombination (red), exciton dissociation (green), and carrier transport (black). The size of the arrows is proportional to the relative amplitude of the depicted process.}\label{fig5}
\end{figure}

The fits give an exciton lifetime of 2ps, exciton dynamics of 30 and 600fs, and hole dynamics of 9 and 260ps (Table \ref{tab1}). There exists a significant body of work on CNT exciton dynamics that we draw upon to provide a physical interpretation of these timescales (graphically illustrated in Fig. \ref{fig5}).\cite{Flach_2020,Mehlenbacher_2013,Mehlenbacher_2016,Grechko_2014,Birkmeier_2022,Crochet_2012,Mehlenbacher_2015,Luer_2008} First, the 2ps exciton lifetime is typical for films made from CNTs of similar lengths and defect density.\cite{Flach_2020,Mehlenbacher_2015} While diffusing along CNTs (at $\sim$500nm$^2$/ps), excitons encounter defects in the tube walls and ends causing recombination.\cite{Birkmeier_2022,Flach_2020} Indeed, 500nm$^2$/ps is so fast that the exciton lifetime is mostly set by the distance between defects. Second, we assign the 30 and 600fs timescales to dissociation of excitons to charges by C\textsubscript{60}. These timescales are not observed in the (6,5) film and so must be caused by exciton dissociation. Based on the diffusion constant, excitons can travel at most 4nm in 30fs, so these excitons are formed adjacent to or nearby C\textsubscript{60}. Third, we assign the 600fs timescale to excitons that must appreciably diffuse or first transfer to another CNT. Regarding the 600fs timescale, we know from prior work that energy transfer between CNT occurs at intersections of crossing tubes.\cite{Mehlenbacher_2015,Mehlenbacher_2013} At the densities of these films, it takes an exciton an average of 500fs to diffuse to a suitable intersection.\cite{Mehlenbacher_2015,Grechko_2014} Moreover, we know for certain that exciton transfer prior to dissociation contributes to photocurrent because the cross-peaks in Fig. \ref{fig3}b exhibited a similar 500fs timescale. Therefore, we assign the 600fs timescale to excitons that must appreciably diffuse or first transfer to another CNT before dissociating.  We note that the (6,5)/(7,5) cross-peak data also had a 30fs timescale, meaning photocurrent is generated both by excitons immediately transferring between parallel tubes and by excitons first diffusing to crossing points.\cite{Mehlenbacher_2016} 

\par

 As established by our experiments, the 9 and 260ps timescales are associated with holes. Less is known about hole transfer in our CNT devices. We tentatively assign the two timescales to fast charge buildup at the anode and slower transport through the CNT layer (Fig. \ref{fig5}). Hole dynamics do not impact efficiency in these devices because charge transfer excitons do not form between CNT and C\textsubscript{60}, and all charges will inevitably be collected.\cite{Classen_2018} Therefore, we leave the precise interpretation of these timescales for a future study. 

\bigskip
\section{Summary and Implications for Device Design}\label{sec4}

The photocurrent experiments presented here determine that 89\% of the current derives from excitons that are harvested within 30fs of their generation (Table \ref{tab1}; row 3).  These excitons are generated adjacent to or nearby C\textsubscript{60}. Only 11\% of the photocurrent is created by excitons that find the C\textsubscript{60} interface within 600fs via diffusion and hopping. Indeed, after about 1ps, excitons continue to transfer between CNT but no longer generate current. The exciton lifetime of 2ps is much longer than these timescales. Thus, even though excitons continue to exist in the films for extended periods of time, most do not contribute to the current. 

\par

This new information dramatically impacts our thinking for how best to design CNT photovoltaics. Based on transient absorption spectroscopy of films, our efforts have been focused on improving exciton transfer. Towards that end, film morphologies were tested that changed the distance between tube intersections, causing a corresponding change in exciton transfer.\cite{Mehlenbacher_2013,Grechko_2014} Exciton transfer between CNTs was improved 10-fold by removing residual polymer wrapping, enabling more direct CNT contact.\cite{Mehlenbacher_2016} Defect densities were modified to extend exciton lifetimes.\cite{Flach_2020} These experiments led to a better understanding of excitons  in these films, but did not improve device efficiency. It now makes sense that the above efforts did not improve efficiency, knowing that the vast majority of current is generated in 30fs. An alternative direction would be to use microstructures with C\textsubscript{60} intercalated into the CNT layer, thereby leveraging the 30fs process. In our current films, CNTs bundle into fibrils of $\sim$7-19 tubes.\cite{Grechko_2014} Thus, disaggregating bundles may help excitons reach C\textsubscript{60}. 

\par

The default approach for assessing exciton dynamics is transient absorption of thin films.  That approach relies on the assumption that the exciton dynamics deduced in films are reflected in the current-generating pathways of the corresponding device. As we show here, that assumption is a poor approximation for CNT photovoltaics. It will be interesting to apply this new spectrometer design to other photovoltaic materials to assess whether the conclusions drawn from transient absorption hold true in those working devices. 

\bigskip
\section{Methods}\label{sec5}

\bigskip
\subsection{Spectrometer}\label{subsec5}

The pump-probe and 2D spectrometer (Fig. \ref{fig1})  that can simultaneously measure photoabsorption and photocurrent with no incoherent background has been described in detail elsewhere.\cite{Faitz_2024} It consists of an Ytterbium laser operating at 100 kHz repetition rate that pumps an optical parametric amplifier to make 800nm pulses, followed by white light (850 – 1300nm) generation using YAG supercontinuum generation.\cite{Dubietis_2017} A pair of pump pulses separated by a delay t\textsubscript{1} is created using a TWINS interferometer configured to create orthogonally polarized pulses oriented at 0\textdegree and 90\textdegree in the lab frame.\cite{Brida_2012} A pair of probe pulses separated by a delay t\textsubscript{3} are created using a transverse, AOM-based pulse shaper, allowing shot-to-shot modulation of time delays and phases at the 100kHz repetition rate.\cite{Kearns_2017,Jones_2019,Shim_2009} A waveplate and a polarizer are placed after the pulse shaper to ensure the probe pulses are oriented at 45\textdegree. A motorized delay stage sets the relative delay between the two sets of pulse pairs, t\textsubscript{2}. With this configuration, four pulses with variable delays (t\textsubscript{1}, t\textsubscript{2}, and t\textsubscript{3}) are created with pulse polarizations of $\left<0\degree,90\degree,45\degree,45\degree\right>$ that eliminates the incoherent mixing background in these samples.\cite{Faitz_2024} The white-light has low fluence ($\sim$200pJ pulses with a $\sim$150$\mu$m spot size at the sample) and the signal is linear with respect to pulse power (see SI). 

\par

When acquiring 2D spectra, the signal intensity is measured as a function of both t\textsubscript{1} and t\textsubscript{3} delays, and a 2D Fourier transform is calculated to obtain the pump and probe frequency axes, respectively.\cite{Hamm_2001} In each experiment, we implement a 4-frame phase cycle using the pulse shaper. For pump-probe experiments, we set t\textsubscript{1}=0 so that there is effectively a single pump pulse and collect the signal as a function of the t\textsubscript{3} delay, thereby resolving the probe frequencies with a Fourier transform. For pump-probe measurements, we also implement an additional pseudo phase cycle by modulating the pump pulses’ relative time delay by half of an optical cycle. To measure population dynamics, spectra are collected as a function of t\textsubscript{2} delay. By measuring the probe light reflected off the sample and onto a single pixel detector, we simultaneously acquire nonlinear photoabsorption and photocurrent signals at the same sample position. The thin film measurements are also performed in reflection.

\par

Due to the low white-light fluence, the time-resolution of the spectrometer is obtained using polarization-gated frequency resolved optical gating (PG-FROG) with a separate 800nm pulse (130fs $\pm$2fs).\cite{Delong_1994} The pump and probe pulses are 17 fs $\pm$4fs and 29 fs $\pm$5fs in duration, respectively, assuming Gaussian time-profiles, giving an instrument response time of 33 fs $\pm$6fs.

\bigskip
\subsection{Samples}\label{subsec5}

The 3 devices use a stacked architecture consisting of 40nm of PEDOT:PSS on an ITO substrate (anode) as an electron blocking material, a photoactive CNT layer deposited via drop casting, 90nm of C\textsubscript{60}, 10nm of the hole blocking material BCP, and 120nm of Ag for the cathode. Each is encased using epoxy and a glass coverslip. The three devices differ in the mixture and thicknesses of the CNT layer: a 4nm thick layer composed solely of (6,5) CNTs, a 4nm thick layer composed solely of (7,5) CNTs, and an 8nm thick layer containing equal portions of (6,5) and (7,5) CNTs that are mixed before depositing. We also performed measurements on a 4nm thick film of (6,5) CNTs deposited on ITO via drop casting in an identical manner as the devices (Fig. \ref{fig1}c). The (6,5) and (7,5) CNTs used in these samples were purified using shear force mixing and aromatic polymers, respectively.\cite{Nish_2007,Graf_2016} 

\bigskip

\backmatter

\bmhead{Supplementary information}
Additional kinetcs, 2D spectra, and equations are found in the Supplementary Information (SI)

\bigskip

\bmhead{Acknowledgements}

\bigskip
\ \\
\section*{Declarations}

\begin{itemize}
\item Funding: This work was supported by Air Force Office of Scientific Research grant FA9550-23-1-0181 and the NSF CHE- 2314378.  Zachary M. Faitz work supported by the NSF Graduate Research Fellowship under Grant No. DGE-2137424. 
\item Conflict of interest/Competing interests: Martin Zanni is co-owner of PhaseTech Spectroscopy, Inc., which manufactures pulse shapers and 2D spectrometers similar to those utilized here.
\item Ethics approval and consent to participate: NlA
\item Consent for publication: N/A
\item Data availability: All data is available upon request from the authors
\item Materials availability: N/A
\item Code availability: All data processing codes are available upon request from the authors 
\item Author contribution: N/A
\end{itemize}

\newpage

%%===========================================================================================%%
%% If you are submitting to one of the Nature Portfolio journals, using the eJP submission   %%
%% system, please include the references within the manuscript file itself. You may do this  %%
%% by copying the reference list from your .bbl file, paste it into the main manuscript .tex %%
%% file, and delete the associated \verb+\bibliography+ commands.                            %%
%%===========================================================================================%%

%\bibliography{maxnames=99}{sn-bibliography}% common bib file
\printbibliography

\end{document}